\documentclass[aps,prd,superscriptaddress,showpacs,preprint,amsmath,amssymb]{revtex4}
\usepackage{graphicx, bm}
\usepackage[usenames]{color}

\usepackage{subcaption}
\begin{document}

\draft
\title{Bounds on anomalous quartic $WWZ\gamma$ couplings in $e^-p$ collisions at the FCC-he}

\author{E. Gurkanli\footnote{egurkanli@sinop.edu.tr}}
\affiliation{\small Department of Physics, Sinop University, Turkiye.\\}

\date{\today}


\begin{abstract}
Non-Abelian structure of the Standard Model predicts the self-interactions of gauge bosons triple gauge couplings (TGC) and quartic gauge couplings (QGC). On the other hand, it is also important to determine the deviations from Standard Model (SM) via anomalous triple gauge couplings (aTGC) and anomalous quartic gauge couplings (aQGC) to test the nature of the Standard Model (SM) and to see the effects of new physics arising from the beyond standard model (BSM). In this study, we focus on the  $e^-p \to j Z\gamma \nu_e \to j (l^{+} l^{-})\gamma \nu_e$ process to examine the $WWZ\gamma$ anomalous quartic gauge couplings (aQGC) at the Future Circular Collider-hadron electron (FCC-he) with center-of-mass energy $\sqrt{s}=5.29$ TeV using with a model-independent way in the effective theory approach. A Cut-based method are applied to analyse the signal and relevant SM background. Using with the total cross-sections, we constrained on the anomalous $ f_ {T,i=0,1,2,5,6,7}/\Lambda^4$ couplings arising from dimension-8 operators at $95\%$ Confidence Level (C.L.) for the dilepton decay of the $Z$-boson. The results are composed with integrated luminosities of ${\cal L}=1,2$ and $3$ $\rm ab^{-1}$ under the systematic uncertainties of $\%0$, $\%5$ and $\%10$. In the calculations, we use the form factor to eleminate the effects of the violation of unitarity. With this, we obtained the limits for the energy scale of the form factor  $\Lambda=1$ TeV. In addition to all this, we use the adventages of $ep$ collisions for analysing the process with much cleaner environment and lower background effects comparing with the $pp$ collisions. The obtained sensitivities on $ f_ {T,i=0,1,2,5,6,7}/\Lambda^4$ are comparable with the experimental results and related phenomenological studies in the literature.

\end{abstract}


\vspace{5mm}

\maketitle

\section{Introduction}

 Due to the SM's non-Abelian nature, it estimates the presence of multi-boson interactions as TGC and QGC. At the same time, anomalous TGC (aTGC) and anomalous QGC (aQGC) are deviations from the SM\cite{PRD93-2016,JHEP10-2017,JHEP10-2021,arXiv:2106.11082}. As a result, it is critical to determine both aTGC and aQGC to check the validity of the SM or look for signs for new physics source from BSM. The study is focus on the sensitivity of anomalous $WWZ\gamma$ couplings. For doing this, we employ the effective Lagrangian formalism, which has been used to parameterize the new physics effects coming from BSM in various particle physics processes. This technique enables the parameterization of any novel physics defined with a model-independent manner.

Many measurements and theoretical studies have been made of with present and future colliders based on proton-proton ($pp$), electron-proton ($ep$) and electron-positron ($ee$) collisions \cite{JHEP06-2020,OYULMAZ,JI-CHONG,MARANTIS,SC1,SC2,ATLAS-PRL2015,ATLAS-PRD2016,PLB811-2020,Stirling,Leil,ALEPH-Barate,DELPHI-Abreu,L3-Acciarri,OPAL-Abbiendi,CDF-Gounder,D0-Abbott,CMS-Chatrchyan,ATLAS-Aaboud,LHeC-FCC-he-WWgg-Ari1,LHeC-FCC-he-WWgg-Ari2,Bervan,Chong,Koksal,Stirling1,Atag,Eboli1,Sahin,Koksal1,Chapon,Koksal2,Senol,Koksal3,Yang,Eboli2,Eboli4,Bell,Ahmadov,Schonherr,Wen,Ye,Perez,Sahin1,Senol1,Baldenegro,Fichet,Pierzchala,Gutierrez,Belanger,Aaboud,Eboli,Eboli3,Gutierrez-EPJC81-2021,twiki.cern,Eboli-PRD101-2020}. Because of the intense interactions associated with $pp$ collisions, the LHC is not useful for precise measurements. At this point, an $ep$ collider might be  an alternative to the LHC's physics program. Furthermore, due to the high luminosity and collision energy of $ep$ colliders, the impacts of new physics arising from BSM can be observed by looking into the interaction of the $W$ and $Z$ boson with the $\gamma$, which needs precise measurement of $WWZ\gamma$ couplings. FCC-he will generate $ep$ collisions with a maximum center-of-mass energy  5.29 TeV which is planned to collide electrons with energies ranging from 60 GeV to 140 GeV with protons ranging from 7 TeV to 50 TeV with an integrated luminosities of ${\cal L}=1, 2, 3$ ${\rm ab^{-1}}$ \cite{LHeC-FCChe-2020, FCC-CDR} . In this work, we only used the maximal center-of-mass energy scenerio 5.29 TeV to evaluate $WWZ\gamma$ couplings.

We take the effective Lagrangian technique in our work to see the new physics effects arising from the BSM by adding extra terms to the SM's Lagrangian. In this paper, we give our results for the total cross-section for the process $e^-p \to j Z\gamma \nu_e $  at the FCC-he in a model-independent way. Bounds on the aQGC $WWZ\gamma$ are composed using with $\sqrt{s}$=5.29 TeV and ${\cal L}=1, 2, 3$ ${\rm ab^{-1}}$ for the dilepton final state ($Z\to l^+l^-$).

The paper are composed with the following subsections: In Section II, we define the effective Lagrangian approach. In Section III, we calculate the cross-section for the process $e^-p \to j Z\gamma \nu_e $ with the selected kinematic cuts. In Section IV, we obtain the sensitivity of the anomalous $ f_ {T,i}/\Lambda^4$ couplings at $95\%$ C.L. at the FCC-he. In Section V, we sum up our results.

\section{Effective Theory Approach}

The Effective theory is convenient to see the effects arising from the BSM in a model-independent manner and a good starting point to search the new physics. While doing this, some extra terms are added to SM Lagrangian to compose the EFT lagrangian which are including higher dimension operators and the new physics scale $\Lambda$ that is being identified in the experiments. These operators have been described by either linear or non-linear effective Lagrangians. The effective field theory Lagrangian including the higher-dimensional operators is given below \cite{LHeC-FCC-he-WWgg-Ari1,LHeC-FCC-he-WWgg-Ari2,LHeC-FCC-he-WWgg-Gurkanli,Eboli3,Degrande}:

\begin{equation}
{\cal L}_{EFT}={\cal L}_{SM}+\sum_{i}\frac{c_i^{(6)}}{\Lambda^2}{\cal O}_i^{(6)}
+\sum_{j}\frac{c_j^{(8)}}{\Lambda^4}{\cal O}_j^{(8)}+...,
\end{equation}

\noindent Here, dimension-6 operators gives the major contribution to the multi-boson vertices. These dimension-6 operators can be classified in two sets CP-Conserving and CP-Violating. The CP-conserving operators are 

\begin{eqnarray}
O_{WWW}=Tr[W_{\alpha\beta}W^{\beta\rho}W^{\alpha}_{\rho}],    \\
O_{W}=(D_{\alpha}\Phi)^{\dagger}W^{\alpha\beta}(D_{\beta}\Phi),  \\
O_{B}=(D_{\alpha}\Phi)^{\dagger}B^{\alpha\beta}(D_{\beta}\Phi),   
\end{eqnarray}

while the CP-violating operators are 

\begin{eqnarray}
O_{\tilde{W}WW}=Tr[\tilde{W}_{\alpha\beta}W^{\beta\rho}W^{\alpha}_{\rho}],    \\
O_{\tilde{W}}=(D_{\alpha}\Phi)^{\dagger}\tilde{W}^{\alpha\beta}(D_{\beta}\Phi),   
\end{eqnarray}

where the $\Phi$ shows the Higgs doublet. Here, $D_\mu\Phi=(\partial_\mu + igW^j_\mu \frac{\sigma^j}{2} + \frac{i}{2}g'B_\mu )\Phi$ is the covariant derivates of the Higgs doublet which $W^{\mu\nu}$ and $B^{\mu\nu}$ are the gauge field strength tensors for $SU(2)_L$ and $U(1)_Y$. In Eqs. 2-6, $O_{WWW}$, $O_{W}$ and $O_{\tilde{W}WW}$ effect the triple gauge boson interactions with the quartic ones simultaneously \cite{Eboli3}. In the low energy region, there is no Higgs mechanism in the non-linear representation that the lowest order operators describe dimension-6 operators. However, the SM gauge symmetry in the linear representation is broken by the conventional Higgs mechanism. Consequently, the lowest order operators that define the possible deviations of the QGC from the SM are dimension-8 in linear representation. To separate the effects of the aQGC in our study, we take into account effective operators that lead to pure aQGC without an aTGC. For aQGC, dimension-8 is the lowest order of “purely” quartic interactions. Therefore, genuine quartic vertices are of dimension-8. The 3 type of dimension-8 aQGC operators are defined in Table I. Also, we give the all dimension-8 operators with the corresponding quartic vertices in Table II. Because of the poor sensitivity of the M-type operators, we only focused on T-type operators in this study.

\begin{table}
\caption{Dimension-8 operators related to aQGC. \cite{Eboli3}.}
\begin{center}
\begin{tabular}{|c|c|}
\hline
\multicolumn{2}{|c|}{S-type operators} \\
\hline
\hline
$O_{S, 0}$     & $[(D_\mu\Phi)^\dagger (D_\nu\Phi)]\times [(D^\mu\Phi)^\dagger (D^\nu\Phi)]$     \\
\hline
$O_{S, 1}$     &  $[(D_\mu\Phi)^\dagger (D^\mu\Phi)]\times [(D_\nu\Phi)^\dagger (D^\nu\Phi)]$    \\
\hline
\multicolumn{2}{|c|}{M-type operators}    \\
\hline
\hline
$O_{M, 0}$   & $Tr[W_{\mu\nu} W^{\mu\nu}]\times [(D_\beta\Phi)^\dagger (D^\beta\Phi)]$    \\
\hline
$O_{M, 1}$   & $Tr[W_{\mu\nu} W^{\nu\beta}]\times [(D_\beta\Phi)^\dagger (D^\mu\Phi)]$  \\
\hline
$O_{M, 2}$   & $[B_{\mu\nu} B^{\mu\nu}]\times [(D_\beta\Phi)^\dagger (D^\beta\Phi)]$  \\
\hline
$O_{M, 3}$   & $[B_{\mu\nu} B^{\nu\beta}]\times [(D_\beta\Phi)^\dagger (D^\mu\Phi)]$  \\
\hline
$O_{M, 4}$   & $[(D_\mu\Phi)^\dagger W_{\beta\nu} (D^\mu\Phi)]\times B^{\beta\nu}$  \\
\hline
$O_{M, 5}$   & $[(D_\mu\Phi)^\dagger W_{\beta\nu} (D^\nu\Phi)]\times B^{\beta\mu}$  \\
\hline
$O_{M, 6}$   & $[(D_\mu\Phi)^\dagger W_{\beta\nu} W^{\beta\nu} (D^\mu\Phi)]$  \\
\hline
$O_{M, 7}$   & $[(D_\mu\Phi)^\dagger W_{\beta\nu} W^{\beta\mu} (D^\nu\Phi)]$  \\
\hline
\hline
\multicolumn{2}{|c|}{T-type Operators}\\
\hline
$O_{T_0}$   &  $Tr[W_{\lambda\nu} W^{\lambda\nu}]\times Tr[W_{\mu\rho}W^{\mu\rho}]$  \\
\hline
$O_{T_1}$   &  $Tr[W_{\lambda\rho} W^{\mu\nu}]\times Tr[W_{\mu\nu}W^{\lambda\rho}]$  \\
\hline
$O_{T_2}$   &  $Tr[W_{\mu\rho} W^{\rho\nu}]\times Tr[W_{\nu\lambda}W^{\lambda\mu}]$  \\
\hline
$O_{T_5}$   &  $Tr[W_{\alpha\beta} W^{\alpha\beta}]\times B_{\mu\nu}B^{\mu\nu}$  \\
\hline
$O_{T_6}$   &  $Tr[W_{\alpha\beta} W^{\rho\lambda}]\times B_{\rho\lambda}B^{\alpha\beta}$  \\
\hline
$O_{T_7}$   &  $Tr[W_{\mu\rho} W^{\rho\nu}]\times B_{\nu\lambda}B^{\lambda\mu}$  \\
\hline
\end{tabular}
\end{center}
\end{table}

\begin{table}
\caption{Dimension-8 operators with the related quartic vertex.}
\begin{center}
\begin{tabular}{|l|c|c|c|c|c|c|c|c|c|}
\hline
& $WWWW$ & $WW$ZZ & ZZZZ & $WWZ \gamma$ & $WW\gamma \gamma$ & ZZZ$\gamma$ & ZZ$\gamma \gamma$ & Z$ \gamma\gamma\gamma$ & $\gamma\gamma\gamma\gamma$ \\
\hline
\cline{1-10}
$O_{S_0}$, $O_{S_1}$                     & + & + & + &   &   &   &   &   &    \\
$O_{M_0}$, $O_{M_1}$, $O_{M_6}$, $O_{M_7}$ & + & + & + & + & + & + & + &   &    \\
$O_{M_2}$, $O_{M_3}$, $O_{M_4}$, $O_{M_5}$ &   & + & + & + & + & + & + &   &    \\
$O_{T_0}$, $O_{T_1}$, $O_{T_2}$           & + & + & + & + & + & + & + & + & +  \\
$O_{T_5}$, $O_{T_6}$, $O_{T_7}$           &   & + & + & + & + & + & + & + & +  \\
$O_{T_8}$, $O_{T_9}$                     &   &   & + &   &   & + & + & + & +  \\
\hline
\hline
\end{tabular}
\end{center}
\end{table}

\begin{table}
\caption{Selected kinematical cuts for $WWZ\gamma$ signal at the FCC-he.}
\begin{tabular}{|c|c|c}
\hline
Kinematic cuts & $f_{Ti}/\Lambda^{4}$ \\
\hline
\hline
Cut1   & \, $p^l_T>20$ GeV, $|\eta^{l}| < 2.5$    \\
\hline
Cut2   &  $|\eta^{\gamma}| < 2.5$ , $p^\nu_T>20$ GeV \\
\hline
Cut3  & \multicolumn{1}{|c|}{ $\Delta R(\ell^{+},\ell^{-}) > 0.4$, $\Delta R(\gamma,\ell) > 0.4$ }\\
\hline
Cut4   & 80 GeV $< M_{\ell^+\ell^-} < 100$ GeV \\
\hline
Cut5   & $p^\gamma_T > 400$ GeV\\
\hline
\end{tabular}
\end{table}

\section{A Cut-Based Analysis for the process $e^-p \to j Z \gamma  \nu_{e}$}

In our work, we examine the process $e^-p \to j Z \gamma  \nu_{e}$ and the relevant SM background using the MadGraph5\_aMC@NLO \cite{MadGraph} package. Using FeynRules \cite{AAlloul} package with dimension-8 effective Lagrangians, the $WWZ\gamma$ aQGC are imported to the MadGraph5\_aMC@NLO. Briefly, signal and relevant SM background events for the process $e^-p \to j Z\gamma \nu_e \to j (l^{+} l^{-})\gamma \nu_e$ are produced at tree-level by importing dim-8 aQGC implemented UFO model file into MadGraph5\_aMC@NLO. In the study, a parton-level analysis is presented without taking into account any detector response. The total cross-sections of the process $e^-p \to j Z \gamma  \nu_{e}$ for the dilepton final state including aQGCs represented by the following equation.

\begin{eqnarray}
\sigma_{Tot}\Biggl( \sqrt{s}, \frac{f_{T,i}}{\Lambda^{4}}\Biggr)
&=& \sigma_{SM}( \sqrt{s} )
+\sigma_{Int}\Biggl( \sqrt{s}, \frac{f_{T,i}}{\Lambda^{4}}\Biggr)
+ \sigma_{BSM}\Biggl(\sqrt{s}, \frac{f^2_{T,i}}{\Lambda^{8}},
\frac{f_{T,i}}{\Lambda^{4}} \Biggr), \nonumber\\
&&i= 0-2,5-7  
\end{eqnarray}

Here; $\sigma_{SM}$ is the SM cross-section, $\sigma_{Int}$ is the interference between SM and BSM. Finally, $\sigma_{BSM}$ is the contribution coming from only the BSM. As is known, $p^{\gamma}_T$ is a useful tool to distinguish the signal and background events due to the behaviour of the high dimensional operators at the large $p^{\gamma}_T$ region. For this reason, we gave the $p^{\gamma}_T$ distributions in Fig.3 to compare the and $f_{T,i}/\Lambda^4$ anomalous couplings and the SM backgrounds for dilepton final state. The distributions contains $\frac{f_{T,i}}{\Lambda^4}$ with $i=0,1,2,5,6,7$ and relevant SM background. This analysis could help to determine the optimized value of photon transverse momentum $p^{\gamma}_T$ to seperate the signal and relevant SM background. Apart from this, other chosen kinematic cuts in our calculations are given in a flow in Table III. Accordingly, we consider the  $p^l_T>20$ GeV, $|\eta^{l}| < 2.5$ as CutI and $|\eta^{\gamma}| < 2.5$ , $p^\nu_T>20$ GeV as Cut2. For Cut3, we applied the angular seperations $(\Delta R =( (\Delta \phi)^2+ (\Delta \eta)^2)^{1/2}$) for dileptons and charged lepton-photon as $\Delta R(l^{+}, l^{-}) > 0.4$ , $\Delta R(\gamma,l) > 0.4$. We also used invariant mass cut for final state dilepton $80 \ GeV < M_{l^{+}l^{-}} < 100 \ GeV$ closest the mass of $Z$ boson with tagged Cut4. Finally, we added the photon transverse momentum $p_{T}^{\gamma}>400$ GeV to seperate the signal and the SM background clearly in Cut5. 

 To see the effects of selected kinematical cuts, we give the number of events including aQGCs $f_{T,0,2,5}/\Lambda^4$ and SM background for the process $e^-p \to j Z \gamma  \nu_{e}$ that the $f_{T,0,2,5}/\Lambda^4$ are taken one at a time in Table IV. Also the total cross-sections ($\sigma_{Tot}$) are composed for $Z \to l^{+}l^{-}$ final state at the FCC-he with $\sqrt{s}=5.29$ TeV.

In Figs. 4-5, the lines are show the deviations from the SM as a function of $f_{T, i}/\Lambda^4$. In these figures, we take into account the dilepton decay of the $Z$-boson $Z \to \l^{+} \l^{-}$ for the process $e^-p \to j Z \gamma  \nu_{e}$ with $l^{-}=e^-, \mu^{-}$ , $l^{+}=e^+, \mu^{+}$.

\section{Sensitivities on the aQGC $f_{T,i}/\Lambda^4$ at the FCC-he}

 Starting from the effective Lagrangian which contains the classes of dimension-8 operators for quartic gauge vertex:

\begin{equation}
{\cal L}_{eff}= {\cal L}_{SM} +\sum_{j=0}^{7}\frac{f_{M, j}}{\Lambda^4}O_{M, j}+\sum_{i=0}^{9}\frac{f_{T, i}}{\Lambda^4}O_{T, i}
+\sum_{k=1}^2\frac{f_{S, k}}{\Lambda^4}O_{S, k},
\end{equation}

\noindent where ${\cal L}_{SM}$ is the SM Lagrangian while $O_{M, j}$, $O_{T, i}$ and $O_{S, k}$ are the dimension-8 operators, respectively \cite{Eboli3,Degrande}.

The effects of new physics coming from the $WWZ\gamma$ coupling qualified by 6 anomalous parameters classified in $f_{T,i}/\Lambda^4$ in Eq. (7) can be tested by $\chi^2$ method at the $95\%$ Confidence Level:

\begin{equation}
\chi^2(f_{T,i}/\Lambda^4)=\Biggl(\frac{\sigma_{Tot}(f_{T,i}/\Lambda^4)-\sigma_{SM}}
{\sigma_{SM}\sqrt{(\delta_{st})^2 + (\delta_{sys})^2}}\Biggr)^2.
\end{equation}

\noindent Here, $N_{SM}$ and $\delta_{st}=\frac{1}{\sqrt{N_{SM}}}$ are the number of events related to the SM and the statictical error, respectively. On the other hand, we consider the systematic uncertainties while obtaining the sensitivities. Systematic uncertainties may arise from related backgrounds, jet-photon misidentification, integrated luminosities, photon efficiencies, and lepton identification. Therefore, calculations are composed under the systematic uncertainties with the values of $\delta_{sys}=\%5$ and $\%10$ \cite{JHEP10-2017,JHEP10-2021,ATLAS-PRD2016}.

\begin{equation}
N_{SM}={\cal L}\times \sigma_{SM}.
\end{equation}

Besides, effective theory is problematic in the high energy region and violates unitarity. In theory, the cross-sections increase harshly with the energy, which must be controlled. A practical way to remove the unphysical behavior is to clip beyond the energy where it starts to violate. To do this, we first give the invariant mass distributions of the final state bosons $M_{\gamma Z}$ in Fig.2. As seen in Fig.2,  signals started to increase harshly near 1 TeV compared with the SM background. The theory is starting to violate after 1.0 TeV, and the contributions coming from beyond this region are unphysical. We have chosen the $\Lambda = 1$ TeV to eliminate these contributions and fix the unitarity issue. Here, the FF value is added by a factor while computing the cross-sections with MadGraph5\_aMC@NLO under the unitarization procedure\cite{Eboli4}.

 To handle this situation, we implied the dipole form factor given below to reduce the unphysical behaviour of total cross-sections \cite{Eboli4}.

\begin{equation}
FF=\frac{1}{(1+\frac{\hat s}{\Lambda^2})^2}. \hspace{1cm}
\end{equation}

We give the limits on the aQGC $f_{T,i}/\Lambda^4$ for the process the $e^-p \to j Z \gamma \nu_{e}$ production at the FCC-he to see the restricting effects of future colliders on aQGCs in Tables V and VI. These results are composed via $e^-p \to j Z \gamma \nu_{e}$ process at the FCC-he with maximal $\sqrt{s}= 5.29$ TeV with the total integrated luminosity ${\cal L}=1,2,3$ ${\rm ab^{-1}}$. 

Our best limits ${\cal O}_{T,j}$ are given in Eq. (12-14) for the FCC-he with ${\cal L}=3$ ${\rm ab^{-1}}$ at $95\%$ C.L. with the dilepton final state for $\Lambda =\infty$ and $\delta_{sys}=0\%$ systematic uncertainties:

\begin{eqnarray}
\frac{f_{T0}}{\Lambda^{4}}&=& [-0.90; 0.72] \hspace{1mm} {\rm TeV^{-4}}, \\
\frac{f_{T5}}{\Lambda^{4}}&=& [-0.47; 1.39] \hspace{1mm} {\rm TeV^{-4}}, \\
\frac{f_{T6}}{\Lambda^{4}}&=& [-3.34; 2.88] \hspace{1mm} {\rm TeV^{-4}}.
\end{eqnarray}

As a result, we constrained on the aQGCs $f_{T,i}/\Lambda^{4}$ for the process $e^-p \to j Z \gamma \nu_{e}$ at the FCC-he with $\sqrt{s}$ = 5.29 TeV. The bounds are given with $\Lambda =1$ TeV and $\Lambda =\infty$ under various systematic uncertainties in Table V and VI, respectively.

\begin{table}
\caption{Number of events for the process $e^-p \to j Z \gamma \nu_{e}$ and SM background for each kinematic cut given in Table III.}
\begin{tabular}{|c|c|c|c}
\hline
Kinematic cuts & Signal ($f_{T0}/\Lambda^{4}$ = 7 TeV$^{-4}$) & Standard Model \\
\hline
\hline
Cut0 & 2750 & 2047  \\
\hline
Cut1  & 2750 & 2047 \\
\hline
Cut2  & 2675 & 2000 \\
\hline
Cut3  & 2675 & 2000   \\
\hline
Cut4  & 2509 & 1878   \\
\hline
Cut5  & 516 & 17  \\
\hline
 & Signal($f_{T2}/\Lambda^{4}$ = 7 TeV$^{-4}$) & Standard Model \\
\hline
\hline
Cut0 & 2117 & 2047  \\
\hline
Cut1  & 2117 & 2047    \\
\hline
Cut2  & 2066 & 2000    \\
\hline
Cut3  & 2066 & 2000   \\
\hline
Cut4  & 1940 & 1878 \\
\hline
Cut5  & 54 & 17   \\
\hline
\hline
 & Signal($f_{T5}/\Lambda^{4}$ = 7 TeV$^{-4}$) & Standard Model \\
\hline
\hline
Cut0 & 2570 & 2047  \\
\hline
Cut1  & 2570 & 2047 \\
\hline
Cut2  & 2504 & 2000  \\
\hline
Cut3  & 2504 & 2000  \\
\hline
Cut4  & 2340 & 1878 \\
\hline
Cut5  & 450 & 17  \\
\hline
\end{tabular}
\end{table}


\begin{table}
\caption{Constraints on the aQGC $WWZ\gamma$ at $95\%$ C.L. for the process $e^-p \to j Z \gamma \nu_{e}$ at the FCC-he with $\delta_{sys}=0\%, 5\%, 10\%$ and $\Lambda=1$ TeV are given.}
\begin{tabular}{|c|c|c|c|c|}
\hline
\multicolumn{5}{|c|}{$\Lambda=1$ TeV } \\
\hline
\hline
Couplings (TeV$^{-4}$) & & ${\cal L}=1$ ab$^{-1}$ & ${\cal L}=2$ ab$^{-1}$ & ${\cal L}=3$ ab$^{-1}$ \\
\hline
                      &$\delta=0\%$       &$[-1.16;0.97]$  &$[-0.99;0.81]$  &$[-0.90;0.72]$ \\
$f_{T0}/\Lambda^{4}$  &$\delta=5\%$       &$[-1.19;1.01]$  &$[-1.04;0.86]$  &$[-0.97;0.79]$ \\
                      &$\ \, \delta=10\%$ &$[-1.27; 1.09]$ &$[-1.16;0.98]$  &$[-1.12;0.94]$ \\
\hline
                      &$\delta=0\%$       &$[-3.91;4.26]$  &$[-3.26;3.61]$  &$[-2.93;3.28]$ \\
$f_{T1}/\Lambda^{4}$  &$\delta=5\%$       &$[-4.04;4.38]$  &$[-3.47;3.81]$  &$[-3.20;3.55]$ \\
                      &$\ \, \delta=10\%$ &$[-4.36; 4.71]$ &$[-3.93;4.28]$  &$[-3.76;4.11]$ \\
\hline
                      &$\delta=0\%$       &$[-4.32;3.86]$ &$[-3.67;3.21]$ &$[-3.34;2.88]$  \\
$f_{T2}/\Lambda^{4}$  &$\delta=5\%$       &$[-4.45;3.98]$ &$[-3.88;3.41]$ &$[-3.61;3.15]$  \\
                      &$\ \, \delta=10\%$ &$[-4.78; 4.31]$&$[-4.35;3.88]$ &$[-4.17;3.70]$ \\
\hline
                      &$\delta=0\%$       &$[-0.70;1.62]$  &$[-0.55;1.47]$  &$[-0.47;1.39]$\\
$f_{T5}/\Lambda^{4}$  &$\delta=5\%$       &$[-0.73;1.65]$  &$[-0.60;1.52]$  &$[-0.53;1.45]$\\
                      &$\ \, \delta=10\%$ &$[-0.81; 1.73]$ &$[-0.71;1.63]$  &$[-0.66;1.58]$\\
\hline
                      &$\delta=0\%$       &$[-3.88;4.29]$  &$[-3.23;3.64]$  &$[-2.90;3.31]$ \\
$f_{T6}/\Lambda^{4}$  &$\delta=5\%$       &$[-4.01;4.42]$  &$[-3.44;3.85]$  &$[-3.17;3.58]$ \\
                      &$\ \, \delta=10\%$ &$[-4.33; 4.75]$ &$[-3.91;4.32]$  &$[-3.73;4.14]$ \\
\hline
                      &$\delta=0\%$       &$[-3.66;4.53]$  &$[-3.01;3.89]$  &$[-2.69;3.56]$ \\
$f_{T7}/\Lambda^{4}$  &$\delta=5\%$       &$[-3.78;4.66]$  &$[-3.22;4.09]$  &$[-2.95;3.83]$ \\
                      &$\ \, \delta=10\%$ &$[-4.11; 4.98]$ &$[-3.68;4.56]$  &$[-3.51;4.38]$ \\
\hline
\end{tabular}
\end{table}

\begin{table}
\caption{Same as Table V, but for $\Lambda=\infty$.}
\begin{tabular}{|c|c|c|c|c|}
\hline
\multicolumn{5}{|c|}{$\Lambda=\infty$ } \\
\hline
\hline
Couplings (TeV$^{-4}$) & & ${\cal L}=1$ ab$^{-1}$ & ${\cal L}=2$ ab$^{-1}$ & ${\cal L}=3$ ab$^{-1}$ \\
\hline
                      &$\delta=0\%$       &$[-7.50;8.76]\times10^{-1}$ &$[-6.21;7.47]\times10^{-1}$ &$[-5.56;6.82]\times10^{-1}$ \\
$f_{T0}/\Lambda^{4}$  &$\delta=5\%$                       
                      &$[-7.65;8.92]\times10^{-1}$  &$[-6.47;7.73]\times10^{-1}$ &$[-5.90;7.16]\times10^{-1}$ \\
                      &$\ \, \delta=10\%$ &$[-8.08;9.35]\times10^{-1}$  &$[-7.11;8.38]\times10^{-1}$ &$[-6.69;7.95]\times10^{-1}$ \\
\hline
                      &$\delta=0\%$       &$[-3.03;3.02]$  &$[-2.55;2.53]$ &$[-2.31;2.29]$ \\
$f_{T1}/\Lambda^{4}$  &$\delta=5\%$           
                      &$[-3.09;3.07]$  &$[-2.65;2.63]$ &$[-2.43;2.42]$ \\
                      &$\ \, \delta=10\%$               &$[-3.25; 3.24]$ &$[-2.89;2.87]$ &$[-2.73;2.71]$ \\
\hline
                      &$\delta=0\%$       &$[-3.68;2.59]$  &$[-3.20;2.11]$ &$[-2.96;1.86]$ \\
$f_{T2}/\Lambda^{4}$  &$\delta=5\%$       
                      &$[-3.74;2.65]$  &$[-3.30;2.20]$ &$[-3.08;1.99]$ \\
                      &$\ \, \delta=10\%$ 
                      &$[-3.90; 2.81]$ &$[-3.54;2.45]$
                      &$[-3.38;2.29]$ \\
\hline
                      &$\delta=0\%$       &$[-0.70;0.93]$  &$[-0.57;0.80]$ &$[-0.51;0.74]$ \\
$f_{T5}/\Lambda^{4}$  &$\delta=5\%$       
                      &$[-0.71;0.95]$  &$[-0.60;0.83]$ &$[-0.54;0.77]$ \\
                      &$\ \, \delta=10\%$ 
                      &$[-0.76; 0.99]$ &$[-0.66;0.89]$ &$[-0.62;0.85]$ \\
\hline
                     &$\delta=0\%$       &$[-2.94;3.11]$  &$[-2.46;2.63]$ &$[-2.22;2.38]$ \\
$f_{T6}/\Lambda^{4}$ &$\delta=5\%$       
                     &$[-3.00;3.17]$  &$[-2.56;2.73]$ &$[-2.34;2.51]$ \\
                     &$\ \, \delta=10\%$ 
                     &$[-3.16; 3.33]$ &$[-2.80;2.97]$ &$[-2.64;2.81]$ \\
\hline
                      &$\delta=0\%$       &$[-3.12;3.03]$  &$[-2.63;2.54]$ &$[-2.38;2.30]$ \\
$f_{T7}/\Lambda^{4}$  &$\delta=5\%$       
                      &$[-3.18;3.09]$  &$[-2.73;2.64]$ &$[-2.51;2.43]$ \\
                      &$\ \, \delta=10\%$ 
                      &$[-3.34; 3.26]$ &$[-2.97;2.89]$ &$[-2.81;2.73]$ \\
\hline
\end{tabular}
\end{table}

\section{Conclusions}

In this paper, we have studied the $e^- p \to j Z \gamma \nu_e $ process at the FCC-he for the dilepton decay $Z\to l^{+}l^{-}$ to constrain the anomalous $f_{T,i}/\Lambda^4$ couplings defining dimension-8 effective operators. While doing this, we performed the cross-section for the process $e^- p \to j Z \gamma \nu_e $ with the selected kinematical cuts for the dilepton final state which helps to reduce the relevant SM background. The total cross-sections are calculated including the interference term and the pure new physics term for each anomalous couplings at a time. In addition, $ep$ colliders have a clean experimental environment that is also enable to reach the better results. In the study, we used form factor to handle the unitarity violation at the high energy region. The obtained limits for the anomalous $f_{T,i}/\Lambda^4$ couplings are given in Tables V-VI at $95\%$ C. L. with integrated luminosities of ${\cal L}=1,2,3$ ${\rm ab^{-1}}$ and various systematic uncertainties for dilepton final state.

As can be seen in Tables V and VI, our results are order of $10^{0}$ and  $10^{-1}$ at $\sqrt{s}=$ 5.29 TeV. Many phenomenological and experimental studies are available which has constrained the $f_{T,i}/\Lambda^4$ anomalous couplings in the literature. Besides pp and lepton colliders, there are some phenomenological studies at $ep$ colliders. In \cite{LHeC-FCC-he-WWgg-Ari1} $ep \to \nu_{e} \gamma \gamma j$ , in \cite{LHeC-FCC-he-WWgg-Ari2} $e^{-}p \to e^{-}\gamma^{*}\gamma^{*}p \to e^{-}W^{+}W^{-}p$ and in \cite{LHeC-FCC-he-WWgg-Gurkanli} $e^{-}\gamma^{*}p \to eW\gamma q^{'}X$ processes are stuied. Our results are comparable with the obtained sensitivities in these studies. An experimental study given in \cite{JHEP06-2020}, $z\gamma jj$ final state are studied with the pp collision at the LHC. Also obtained limits are comparable with our results. Apart from all these experimental and phenomenological studies $ep$ colliders are useful for analysing the process with clean environment and lower background effects comparing with $pp$ colliders. There are also several phenomenological and experimental studies in the literature. The limit values in these studies are also generally order of $10^{0}$ and $10^{-1}$. Our results are also compitable with the results in these studies \cite{JHEP06-2020,OYULMAZ,JI-CHONG,MARANTIS,SC1,SC2,ATLAS-PRL2015,ATLAS-PRD2016,PLB811-2020,Stirling,Leil,ALEPH-Barate,DELPHI-Abreu,L3-Acciarri,OPAL-Abbiendi,CDF-Gounder,D0-Abbott,CMS-Chatrchyan,ATLAS-Aaboud,LHeC-FCC-he-WWgg-Ari1,LHeC-FCC-he-WWgg-Ari2,Bervan,Chong,Koksal,Stirling1,Atag,Eboli1,Sahin,Koksal1,Chapon,Koksal2,Senol,Koksal3,Yang,Eboli2,Eboli4,Bell,Ahmadov,Schonherr,Wen,Ye,Perez,Sahin1,Senol1,Baldenegro,Fichet,Pierzchala,Gutierrez,Belanger,Aaboud,Eboli,Eboli3,Gutierrez-EPJC81-2021,twiki.cern,Eboli-PRD101-2020,LHeC-FCC-he-WWgg-Gurkanli}. However, it may be instructive for the future researches related with the  $WWZ\gamma$ aQGCs at FCC-he at the CERN.

\vspace{1cm}

\newpage
\begin{center}
{\bf Acknowledgements}
\end{center}

The numerical calculations reported in this paper were fully performed 
at TUBITAK ULAKBIM, High Performance and Grid Computing Center (TRUBA resources).


\newpage

\begin{figure}[t]
\centerline{\scalebox{1.0}{\includegraphics{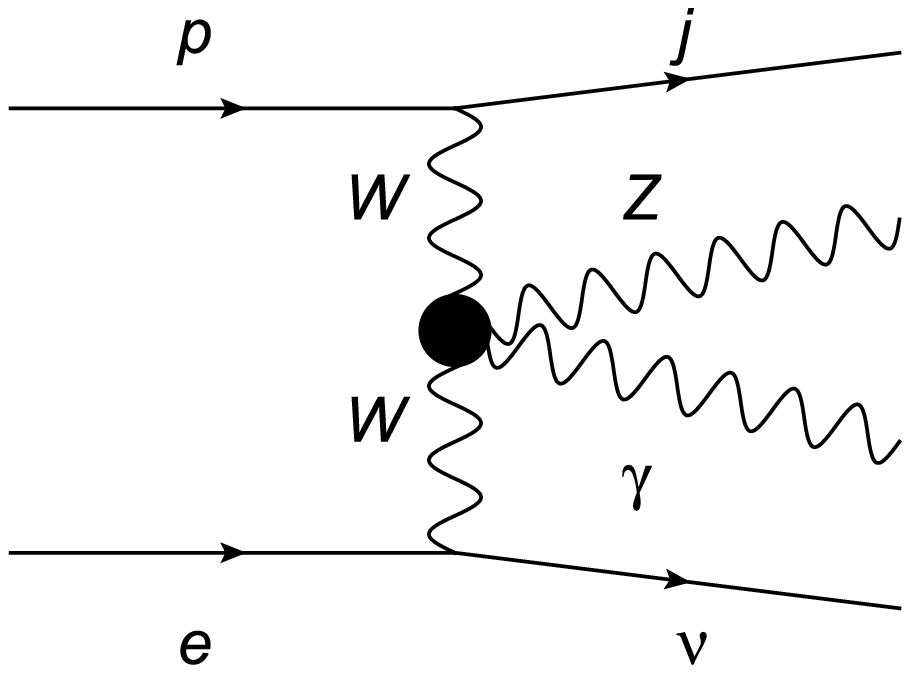}}}
\caption{ \label{fig:gamma1} Diagram for the process $e^-p \to j Z\gamma \nu_e $. New physics effects illustrated by black circle can characterize the quartic gauge couplings.}
\label{Fig.1}
\end{figure}

\begin{figure}[t]
\centerline{\scalebox{0.9}{\includegraphics{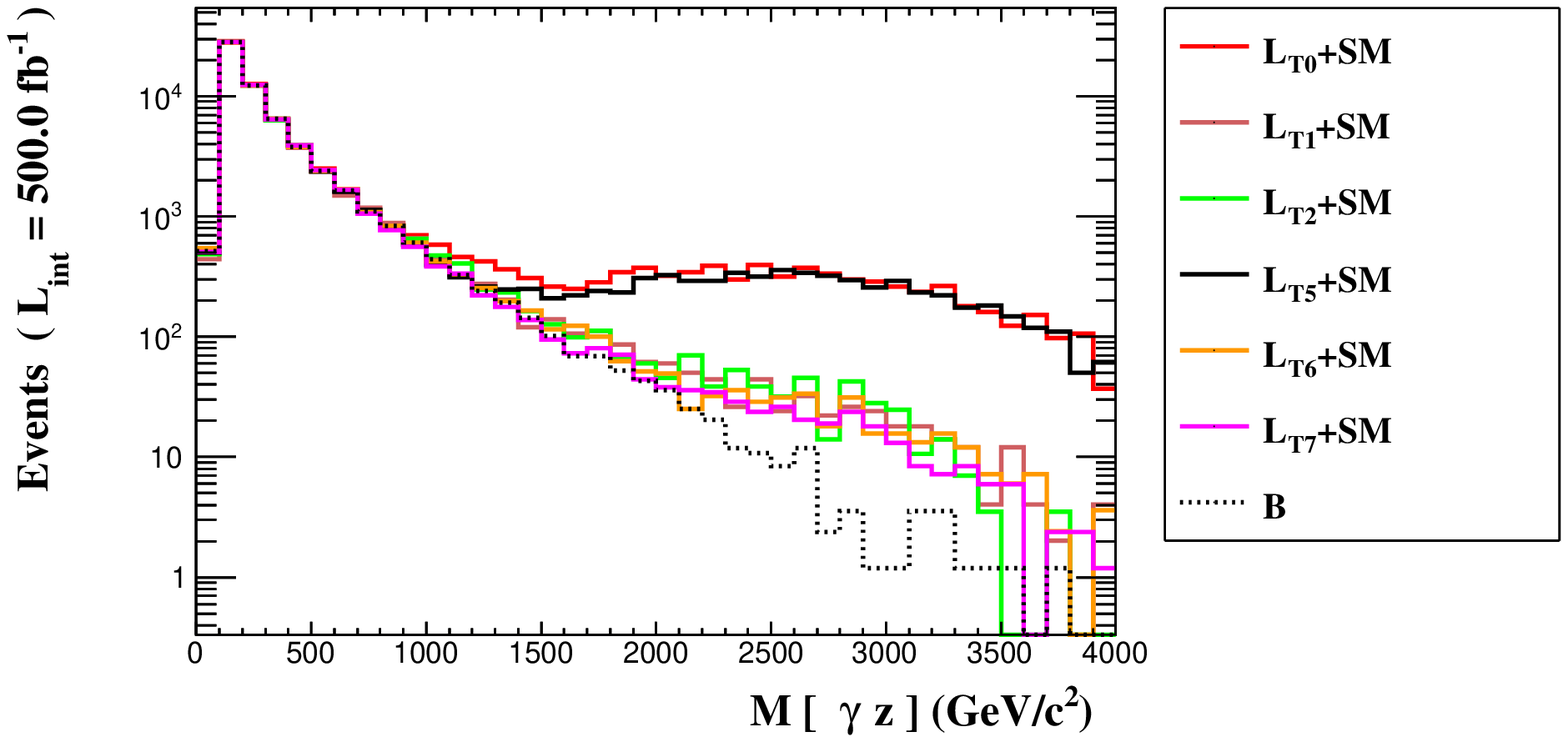}}}
\caption{ \label{fig:gamma2} The number of events as a function of $M_{\gamma Z}$ for the $e^-p \to j Z\gamma \nu_e $ signal and SM background at the FCC-he with $\sqrt{s}=5.29$ TeV. The distributions are for $f_{T,i}/\Lambda^4$  with $i=0,1,2,5,6,7$.}
\label{Fig.2}
\end{figure}

\begin{figure}[t]
\centerline{\scalebox{0.9}{\includegraphics{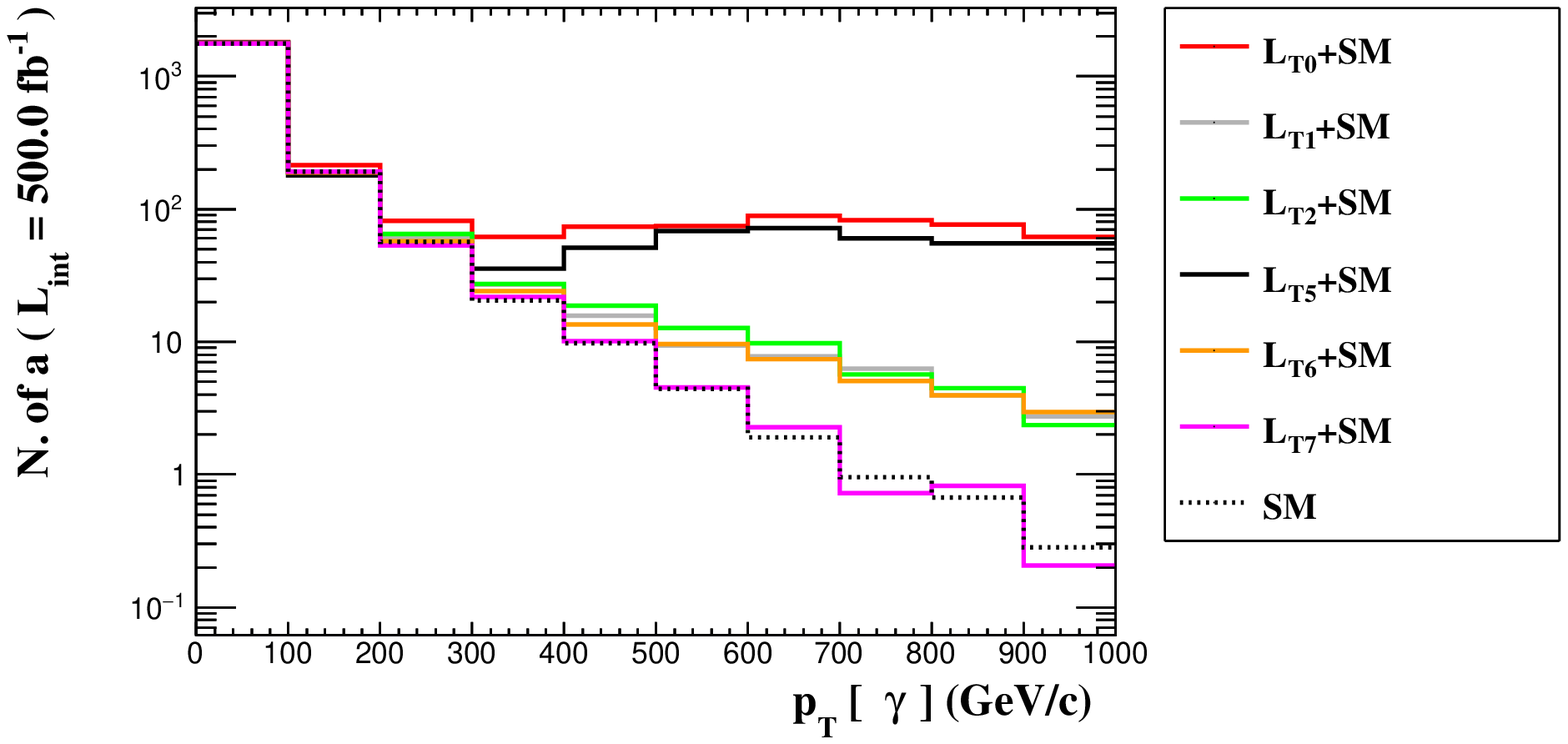}}}
\caption{ \label{fig:gamma2} The number of events as a function of $p^{\gamma}_T$ for the $e^-p \to j Z\gamma \nu_e $ signal and SM background at the FCC-he with $\sqrt{s}=5.29$ TeV. The distributions are for $f_{T,i}/\Lambda^4$  with $i=0,1,2,5,6,7$.}
\label{Fig.2}
\end{figure}

\begin{figure}[t]
\centerline{\scalebox{1.2}{\includegraphics{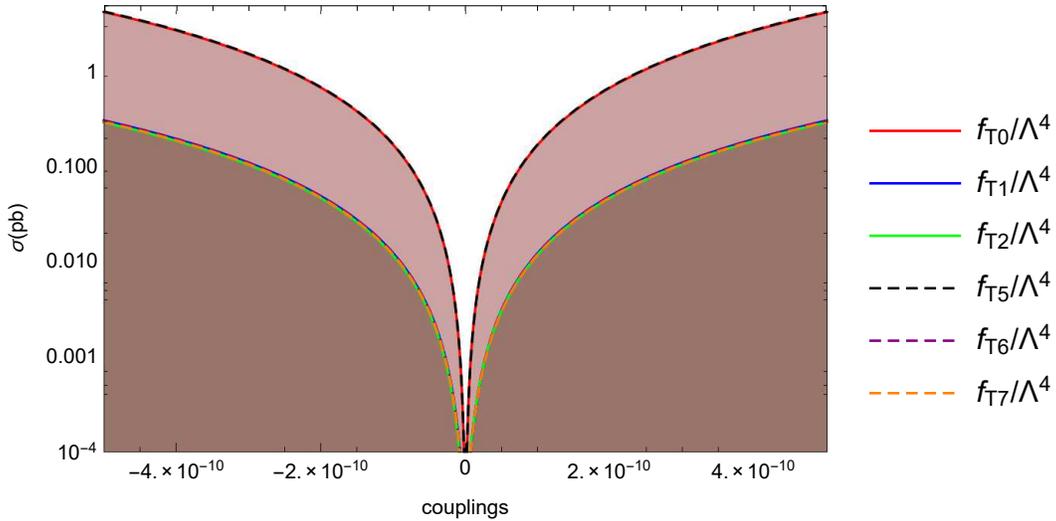}}}
\caption{ \label{fig:gamma2} The total cross-sections of the process $e^-p \to j Z\gamma \nu_e $ as a function of the anomalous couplings $f_{T,i}/\Lambda^4$ for $\Lambda= \infty$ at the FCC-he with $\sqrt{s}=5.29$ TeV.}
\label{Fig.2}
\end{figure}

\begin{figure}[t]
\centerline{\scalebox{1.2}{\includegraphics{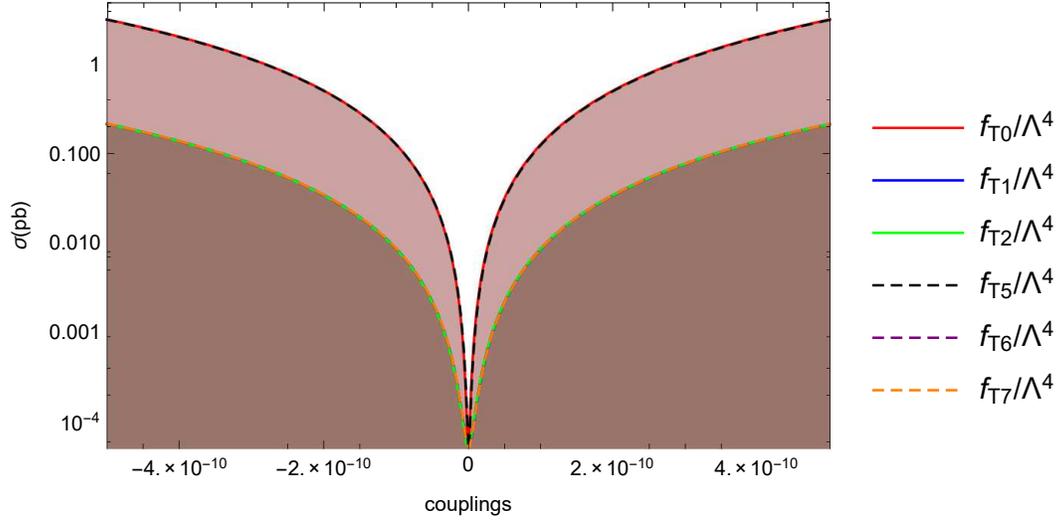}}}
\caption{ \label{fig:gamma1} Same as Fig.4 but for $\Lambda= 1$ TeV }
\label{Fig.1}
\end{figure}

\end{document}